\documentstyle{mn}
\input{epsf}
\newif\ifAMStwofonts 

\def\lesssim{\mathrel{\hbox{\rlap{\hbox{\lower4pt\hbox{$\sim$}}}\hbox{$<$}}}}

\def\gtrsim{\mathrel{\hbox{\rlap{\hbox{\lower4pt\hbox{$\sim$}}}\hbox{$>$}}}}

\def\teff{$T_{\rm eff}$~}

\def\ll_lsun{$\log{\rm (L/L_{\odot})}$~}

\def\masa_msun{$M/ \rm M_{\odot}$~}

\def\m_mstar{$M/M_{*}$~}

\def\mh{${M_{\rm H}/M_*} $ }

\def\lmh{ $\log{{(M_{\rm H}/M_*)}}$ }

\title[Mode trapping in DA White Dwarfs]  {On mode trapping in
pulsating DA white dwarf stars}

\author[O.  G. Benvenuto, A.  H. C\'orsico,  L. G.   Althaus and A. M.
Serenelli] 
{O.   G.   Benvenuto\thanks{Member  of  the  Carrera  del  Investigador
Cient\'{\i}fico, Comisi\'on de  Investigaciones Cient\'{\i}ficas de la
Provincia de Buenos Aires, Argentina.},
A.  H.  C\'orsico\thanks{Fellow of the Consejo Nacional de
Investigaciones     Cient\'{\i}ficas    y     T\'ecnicas    (CONICET),
Argentina.},
L.   G.    Althaus\thanks{Member  of  the   Carrera  del  Investigador
Cient\'{\i}fico y Tecnol\'ogico, CONICET.} and 
A.  M.  Serenelli\thanks{Fellow of  CONICET.}  \\ Facultad de Ciencias
Astron\'omicas  y Geof\'{\i}sicas, Universidad  Nacional de  La Plata,
Paseo del Bosque S/N, (1900) La Plata, Argentina.\\ 
Instituto de Astrof\'{\i}sica de La Plata, IALP, CONICET \\
Emails: obenvenuto,acorsico,althaus,serenell@fcaglp.fcaglp.unlp.edu.ar}

\pagerange{\pageref{firstpage}--\pageref{lastpage}}

\pubyear{2001}

\begin{document}

\maketitle

\label{firstpage}

\begin{abstract} 
The  present  work   is  designed  to  explore  the   effects  of  the
time-dependent element diffusion on the mode trapping properties of DA
white dwarf  models with various  thickness of the  hydrogen envelope.
Our predictions are compared with the standard assumption of diffusive
equilibrium in the trace  element approximation.  We find that element
diffusion markedly weakens the presence of mode trapping originated in
the outer  layers of the  models, even for  the case of  thin hydrogen
envelopes.
\end{abstract}


\section{Introduction} \label{sec:intro}

ZZ  Ceti  or DAV  stars  are variable  white  dwarf  (WD) stars  with
hydrogen-rich  envelopes   (DA  WDs)  that   become  unstable  against
non-radial  $g$(gravity)-modes  when they  move  across the  effective
temperature  ($T_{\rm  eff}$)  range  12500  K  $\gtrsim  T_{\rm  eff}
\gtrsim$ 10700  K. Apart  from the fact  that these cool  WDs pulsate,
they are completely normal DA WDs.  ZZ Ceti stars exhibit light curves
having periodicities in the range  of periods $100\ {\rm s} \lesssim P
\lesssim 1200$  s, with  low amplitudes ($\lesssim  0.30$ magnitudes).
Because the effects of geometric cancellation on the stellar disk make
brightness variations diminish for  progressively higher values of the
harmonic degree ($\ell$), all of the observed periods in ZZ Ceti stars
correspond to  $\ell \leq$  3 (Dziembowski 1977),  being most  of them
dipolar ($\ell$= 1) modes.  The classical $\kappa-\gamma$ mechanism of
overstability acting in the outer  layers of hydrogen has been usually
invoked  as responsible  for  $g$-modes  in ZZ  Ceti  stars (Dolez  \&
Vauclair 1981; Winget et  al.  1982).  More recently, Brickhill (1991)
and  Goldreich \&  Wu (1999)  have proposed  the  ``convective driving
mechanism'' as being the cause of onset of pulsations in such stars.

The  existence  of  the  DAV instability  strip\footnote{Other  hotter
instability strips corresponding to  degenerate pulsators are those of
the  DBVs and  the DOVs  (and PNNVs)  stars, which  correspond  to the
$T_{\rm eff}$ ranges 24000 K $\gtrsim T_{\rm eff} \gtrsim$ 22000 K and
170000  K  $\gtrsim  T_{\rm  eff}  \gtrsim$  80000  K,  respectively.}
provides  astronomers with  an  unique opportunity  to  peer into  the
internal structure  of WDs, which  otherwise would remain  hidden from
observers.   In this  regard, powerful  tools have  been  developed in
recent  years  taking advantage  both  of  important  advances in  the
observational  techniques (Whole  Earth Telescope;  see Nather  et al.
1990) and  the WD  modeling.   As a  result,  valuable constraints  on
fundamental quantities  such as stellar mass,  chemical composition of
the core  and helium and  hydrogen envelope masses have  been assessed
for several  DA WDs  (see, e.g. Pfeiffer  et al.  1996;  Bradley 1998,
2001; Benvenuto et al. 2002).

In spite of the notable success reached in asteroseismological studies
of ZZ  Ceti stars,  some problems in  this field remain  yet unsolved.
One of  these problems is to find  the reason of why  only some modes,
amongst all  the available ones  as predicted by  theoretical studies,
are observed in real stars.  In this regard, notable examples are some
DAVs  with  very  few  independent periodicities  in  their  intrinsic
luminosity variations: G226-29 shows only one period (109.3 s; Kepler,
Robinson \&  Nather 1983); R548 exhibits  two periods (213  and 274 s;
Stover et al. 1980); G117-B15A and  GD 154 pulsate in three periods of
215.2, 271  and 304.4 s (Kepler et  al.  1982), and 402.6  s, 1088.6 s
and 1186.5 s (Pfeiffer  et al.  1996), respectively.  Alarmingly, tens
of excited eigenmodes are predicted by current non-adiabatic pulsation
calculations  for  DA  WD  models  having  structural  characteristics
corresponding to  the above-mentioned  stars.  In this  connection, it
has  been  long  suspected  that  some  efficient  mechanism  of  mode
filtering must be working in  such stars.  By means of this mechanism,
only some few modes would reach large enough amplitudes for them to be
preferentially detected in the light curves.

In a pioneer work, Winget, Van Horn \& Hansen (1981) showed that there
are certain modes that are  ``trapped'' in the outer hydrogen envelope
of stratified  WDs.  These authors found  that a mode  is trapped when
the  local wavelength  (or  an integer  number  of it)  of its  radial
eigenfunction nearly  matches the thickness of  the hydrogen envelope.
Such a mode suffers from an efficient reflection at the density change
across  the hydrogen-helium  chemical  transition.  The  mode is  thus
forced to oscillate with appreciable amplitude only in the low-density
region between the  stellar surface and that interface.   As a result,
the  amplitude  of the  trapped  mode  is  notably reduced  below  the
chemical transition region\footnote{This  picture becomes more complex
when transition  regions other than the  hydrogen-helium interface are
present  in  the  interior  of  star.  However,  their  mode  trapping
capability  is  markedly  weaker  than  that  of  the  hydrogen-helium
transition  (see   Brassard  et  al.   1992ab   and  Bradley  1996).}.
Consequently,  trapped  modes  are  characterized by  low  oscillation
kinetic  energies ($E_{\rm  kin}$) and  therefore large  linear growth
rates. In particular, trapped modes  with periods close to the thermal
time-scale at  the base of the  driving region would  have the largest
amplitudes and they would be the preferentially observed ones (Bradley
\& Winget 1994). This scenario,  which has had great acceptance in the
WD pulsation community, may help explain in a natural way the scarcity
of eigenperiods in the light curves of some ZZ Ceti stars.

Observationally,  signatures  of mode  trapping  have  been found  for
various DAVs  on the basis of  the measured period  ratios between the
dominant  modes,  in  agreement  with  the  predictions  of  adiabatic
numerical calculations  (see Bradley \&  Winget 1991, Brassard  et al.
1992a).  However, recent  studies cast some doubts on  the validity of
the mode  trapping hypothesis.   In particular, seismological  fits to
the  pulsating WD  G117-B15A (Bradley  1998, Benvenuto  et  al.  2002)
indicate that  the observed  period of 215.2  s, which has  the larger
amplitude in the Fourier spectrum,  does not correspond to the trapped
mode predicted by the best fitting WD model.

A  key ingredient  entering  the mode  trapping  studies concerns  the
modeling of  the chemical transition  zones.  In this regard,  we note
that  in  the vast  majority  of  DA  WD pulsation  studies  (Tassoul,
Fontaine \&  Winget 1990;  Brassard et al.   1991, 1992ab;  Bradley \&
Winget  1991;   Bradley  1996,  1998,  2001),  the   modeling  of  the
hydrogen-helium  interface (the most  relevant one  in DA  white dwarf
pulsations) is  performed assuming that the abundance  profiles in the
transition regions  are specified by  the condition of  {\it diffusive
equilibrium}\footnote{On the  opposite point of view, in  the study of
Winget, van Horn \& Hansen  (1981) the chemical interfaces are treated
as {\it true  discontinuities}.}, and in most cases  the trace element
approximation is employed (see  Tassoul et al.  1990).  An improvement
to this  situation is the  work of C\'orsico  et al.  (2001)  who have
carried out  an adiabatic pulsational  analysis of DA  WD evolutionary
models  computed in  a  self-consistent way  with  the predictions  of
time-dependent element diffusion, nuclear  burning, and the history of
the WD (Althaus et al. 2002).   As a result of this study, the authors
found  that, at  least for  massive  hydrogen envelopes  and for  long
periods,  the  effects  of  mode  trapping  in  DA  WDs  are  strongly
diminished as compared  with the situation in which  the trace element
approximation is employed.

As  well  known, the  theory  of  stellar  evolution predicts  massive
hydrogen envelopes  for DA  WDs (Iben \&  MacDonald 1986;  D'Antona \&
Mazzitelli 1991; Althaus et  al. 2002), in agreement with expectations
from some  seismological studies  (Clemens 1994; Bradley  1998, 2001).
However, the  possible existence of DAVs with  thin hydrogen envelopes
has  been put  forward  in some  cases, such  as  the ZZ  Ceti GD  154
(Pfeiffer et al.  1996).  In this regard, we judge it to be worthwhile
to extend our  exploration of the effects of  element diffusion on the
trapping properties  made in  C\'orsico et al.  (2001) to the  case of
less  massive  hydrogen  envelopes.    In  contrast  to  C\'orsico  et
al. (2001), we do not attempt  here a detailed treatment of the pre-WD
evolution. Rather,  we obtain  our initial models  on the basis  of an
artificial  evolutionary procedure.  Another  aim of  this work  is to
compare our  results with the predictions of  equilibrium diffusion in
the trace element approximation.

The paper is  organised as follows.  In Section  2 we briefly describe
our evolutionary-pulsational code and starting models. In Section 3 we
present in detail the pulsational results. Attention is focused mainly
on  the  oscillation kinetic  energies  and  period spacing  diagrams.
Finally, Section 4 is devoted to summarizing our results.

\section{Details of the Computations} \label{sec:comput}

For this work we have  employed the same pulsational code and physical 
ingredients
as in C\'orsico  \& Benvenuto (2002) and Benvenuto  et al. (2002). One
of the  most important aspects of  this study concerns  the modeling of
the chemical  abundance distribution. Specifically,  we are interested
in assessing the  effect of diffusion of nuclear  species on the shape
of the profile of the  chemical composition at the chemical transition
zones.  To this end, we  have included in our calculations the various
processes   responsible   for   element  diffusion.   In   particular,
gravitational settling, chemical and thermal diffusion have been taken
into account (Althaus \& Benvenuto 2000) following the formulation given
by Burgers (1969) for multicomponent plasmas. In our work, the WD
evolution is followed in a self-consistent way with the evolution of
the chemical abundance distribution as given by 
time-dependent element diffusion. In this way, we are avoiding the use
of  the equilibrium diffusion in the trace  element approximation  
usually assumed  in most of white dwarf pulsation studies.  In this 
approximation,
the  transition zone is separated into two parts:
an  upper one  in  which one  element  is dominant  and  the other  is
considered as  a trace  and a lower  region in  which the role  of the
respective  elements  is  reversed.    Because  the
solutions are matched for  fulfilling the condition of conservation of
mass of each element, a discontinuity in the derivative occurs just at
the matching point. The resulting profile remains fixed during the
whole evolution except only for small changes induced by modification in the 
ionization state of the plasma.

In our  code, evolutionary and pulsational  calculations are performed
in  an automatic  way.  After  selecting a  starting stellar  model we
choose an interval  in $P$ and \teff.  The  evolutionary code computes
the model cooling until the hot edge of the \teff-interval is reached.
Afterwards, the pulsation code searches  and computes the whole set of
eigenfrequencies within the selected  period window.  From this stage,
the  eigensolution for the  first model  within the  \teff-interval is
employed as approximate  one for the subsequent model  and iterated to
convergence.  This process  is repeated  for all  of  the evolutionary
models  inside the  chosen \teff-interval.  Note that  the  search for
eigenmodes  is   performed  only  once  in   the  whole  computational
procedure.  We refer  the reader to C\'orsico \&  Benvenuto (2002) and
Benvenuto et  al. (2002), and  references therein for  further details
about our  computational strategy.  Let us  quote that most  of our WD
evolutionary  models  have been  divided  in  about 2000  mesh-points,
whereas for mode calculations we employed up to 5000 mesh-points.

As mentioned in the Introduction,  we undertake this project to assess
the  role of  element diffusion  on  the pulsation  properties of  DAV
models  with  different  hydrogen  envelopes.  Because  thin  hydrogen
envelopes are  not predicted by the standard treatment of stellar
evolution, we construct the starter  models for the computation of the
cooling sequences  by means  of an artificial  evolutionary procedure.
Specifically,  we use  the  artificial heating  technique detailed  in
Althaus \&  Benvenuto (2000).   Here, the stellar  model is  forced to
undergo  an  unphysical  evolution  as  a  result  of  introducing  an
artificial energy release which,  after the star become bright enough,
is  switched off  smoothly.  The  star  relaxes then  to the  physical
cooling  branch after few  tens of  models. We  have checked  that the
stellar model  converges not only  to the physical cooling  branch but
{\it simultaneously}  to a correct internal chemical  profile. To this
end, we  employ different  shapes of the  initial hydrogen  and helium
profiles   at   the  hydrogen-helium   interface.    We  found   that,
irrespective  of  the  precise  shape  of   such  initial  chemical
interface, diffusion evolves  it to a well defined  one far before the
star reaches the ZZ Ceti domain. With respect to the hydrogen envelope
mass,  we elect  the following  values: \lmh= $-3.941$\footnote{We find
that models with more  massive hydrogen envelopes suffer from hydrogen
flashes.   So, we  consider  this value  as  the upper  limit for  the
hydrogen content. Interestingly, this  value is in agreement with what
is predicted when an account  is made of the evolutionary stages prior
to  the WD  formation (see  Althaus et  al.  2002).}, $-4.692$, $-5.672$,
$-6.700$ and $-7.349$. Finally,  the innermost chemical composition of our
models is that of Salaris et al. (1997).

\section{Pulsational Results} \label{sec:pulsation}

In  this work  we have  computed adiabatic  non-radial  $g$-modes with
$\ell =$ 1, 2, 3, for periods ranging from 100 to 1000 s. In addition,
we have  obtained the  forward period spacing  $\Delta P_k=  P_{k+1} -
P_k$ (being $k$ the radial order of mode), and the oscillation kinetic
energy  $E_{\rm kin}$  for each  mode, given by:

\begin{equation} \label{eq1}
E_{\rm kin} = \frac{1}{2} (G M_* R_*^2) \omega_k^2 \int_{0}^{1} 
x^2 \rho \left[ x^2 y_1^2 + x^2 \frac{\ell (\ell +1)}{(C_1 \omega_k^2)^2}
y_2^2\right] dx,
\end{equation}

\noindent where $M_*$  and $R_*$ are the stellar  mass and the stellar
radius respectively, $G$ is  the gravitation constant, $C_1= (r/R_*)^3
(M_*/M_r)$  and  $x =  r  / R_*$.   $\omega_k$, the dimensionless 
eigenfrecuency, is  given by  equation
$\omega_k^2 =  \sigma^2_k (G M_* / R_*^3)^{-1}$,  being $\sigma_k$ the
eigenfrecuency ($P_k= 2  \pi / \sigma_k $). The  quantities $y_1$ and $y_2$
are the dimensionless eigenfunctions (see  Unno et al.  1989 for their
definition).

Pulsational  calculations have
been performed on 0.6 $M_{\odot}$  WD evolutionary models in a $T_{\rm
eff}$ range  covering the entire  observed ZZ Ceti  instability strip.
The  Brunt-V\"ais\"al\"a  frequency   ($N$),  a  fundamental  quantity
entering the  non-radial pulsations of  WDs, is obtained  by employing
the  ``modified Ledoux  treatment''  as in  Brassard  et al.   (1991), 
that is

\begin{equation}
N^{2}=  {{g^{2}  \rho}\over{P}} {{\chi_{T}}\over{\chi_{\rho}}}  \left[
\nabla_{\rm ad} - \nabla + B \right]. 
\end{equation}

Here, $\nabla_{\rm ad}$ and $\nabla$ are the adiabatic and 
the actual temperature gradients. The Ledoux term $B$, for the 
case of a multicomponent 
plasma (M-component plasma), is given by

\begin{equation}
B=     -     {{1}\over{\chi_{T}}}    \sum_{i=1}^{M-1}     \chi_{X_{i}}
{{d\ln{X_{i}}}\over{d\ln{P}}},
\end{equation}

\noindent where

\noindent    $$\chi_{\rho}=\left(   {{\partial   \ln{P}}\over{\partial
\ln{\rho}}} \right)_{T,\{X_{i}\}}$$
 
\noindent $$\chi_{T}=\left(  {{\partial \ln{P}}\over{\partial \ln{T}}}
\right)_{\rho,\{X_{i}\}}$$

\noindent   $$\chi_{X_{i}}=\left(   {{\partial   \ln{P}}\over{\partial
\ln{X_{i}}}} \right)_{\rho,T,\{X_{j\neq i}\}}.$$

Before entering the discussion  of the pulsational results, some words
concerning  the  shape  of  our  chemical  profile  and  the  relevant
pulsation quantities are in order.  To this end we show in Fig.  1 the
internal chemical  abundance distribution, in  Fig. 2 the  Ledoux term
($B$) and in Fig. 3 the resulting Brunt-V\"ais\"al\"a frequency for WD
models at $T_{\rm eff }\approx$ 11800  K for all of the considered \mh
values.  We also show in  these figures the predictions resulting from
the   use  of  the   diffusive  equilibrium   in  the   trace  element
approximation  at  the hydrogen-helium  transition  zone (thin  dotted
lines).  Note the smoothness of the composition at the hydrogen-helium
chemical interface  resulting from  the diffusion processes.   This is
true  even for  the case  of the  thickest hydrogen  envelope, because
diffusion is operative at the  large depths reached by such envelopes.
For each considered \mh value, the hydrogen-helium chemical transition
gives rise to a  marked increase in $B$, which in turn is  translated 
as a smooth
bump into  the Brunt-V\"ais\"al\"a frequency.  In  contrast, the trace
element approximation yields a strongly peaked feature in the $B$ and $N^2$
values at
the hydrogen-helium interface.  It  is important to note that, despite
the  fact  that  the  condition  of diffusive  equilibrium  is  nearly
realised at the bottom of  our thinnest hydrogen envelopes, the use of
the trace element approximation is clearly not an appropriate one even
in the  case of such envelopes.  As  a last remark, note  that the $B$
term exhibits  a very  pronounced peak at  deepest layers, which  is a
result of the steep growth of carbon abundance from a central constant
value of $X_{\rm ^{12}C}\approx  0.15$ to $X_{\rm ^{12}C} \approx 0.4$
at $\log(1 - M_r / M_*) \approx -0.45$. Fig. 3 clearly shows that this
peak   is  further  present   in  the   Brunt-V\"ais\"al\"a  frequency
profile. As we  shall see, the presence of  this pronounced feature in
the  Brunt-V\"ais\"al\"a   frequency  at  high   density  regions  has
non-negligible consequences for  oscillation kinetic energy of certain
modes.

Next, we describe  the results of our pulsation  calculations. To this
end, we  concentrate ourselves on  the $\log (E_{\rm  kin})$-$P_k$ and
$\Delta  P_k$-$P_k$  diagrams.  As   well  known,  the  $\log  (E_{\rm
kin})$-$P_k$  relation enables  us to  infer the  presence  of trapped
modes in the outer layers.  In particular, trapped modes correspond to
local mimima in the $E_{\rm kin}$ spectrum.  In addition, the presence
of mode trapping  is also inferred through the  analysis of the period
spacing  diagrams  (Brassard et  al.   1992ab).   Here, mode  trapping
manifests itself in departures from uniformity in the period spacing of
consecutive overtones. Specifically, trapped modes correspond to local
minima  in $\Delta  P_k $.   This quantitiy  is potentially  useful to
detect  trapped   modes  in  the   light  curves  of   pulsating  WDs.
Nonetheless,  the evaluation  of $\Delta  P_k$ requires  at  least two
periods with consecutive  $k$ values to be present  in the pulsational
pattern;  otherwise this  quantity remains  undefined.  Unfortunately,
the DAVs show  periods with sparsed $k$ values,  making the employment
of $\Delta P_k$ almost useless  to infer the presence of mode trapping
from observations.

Figs.  4 and  5  depict  the $\log  (E_{\rm  kin})$-$P_k$ and  $\Delta
P_k$-$P_k$  diagrams corresponding  to the  \mh values  of  $-3.941$ and
$-7.349$  for $\ell=$  1,  2, 3  modes  for WD  models  at $T_{\rm  eff}
\approx$  11800  K.  The  predictions  of  the time-dependent  element
diffusion  are  shown by  filled  symbols  and  solid lines,  and  the
expectations from the standard  assumption of diffusive equilibrium in
the trace element approximation are  denoted by empty symbols and thin
lines.  In the  interests of
clarity, the scale for the kinetic energy in the case of trace element
approximation is displaced  upwards by 1 dex.  For  short periods, the
oscillation  kinetic  energy  and  period  spacing  distributions  are
qualitatively  similar  for both  treatments  of  diffusion.  This  is
notable  in the  case of  the thin  \mh value.   The outstanding  feature
illustrated by these figures is the fact that from a sufficiently high
radial order, say $k$', and depending on the $\ell$ value, the kinetic
energy  distribution is  quite even  when account  is made  of stellar
models with  diffusively evolving chemical compositions.   This is true
even for the thinnest hydrogen envelopes we have analyzed.  Note that,
particularly  for  the thin hydrogen  envelope,  the  $k$' value  becomes
smaller as  $\ell$ value is  increased. As a  last remark, we  want to
comment  on the  fact that  for  the thickest  hydrogen envelope  (see
Fig. 4), the uniformity of  the kinetic energy distribution (i.e.  the
absence  of  appreciable mode  trapping  signatures)  takes place  for
periods longer than  $\approx$ 400 - 500 s  irrespective of the $\ell$
value\footnote{ A  similar result has  been found by C\'orsico  et al.
(2001) in the  frame of full evolutionary calculations  that take into
account the history of the WD progenitor.}.

The behaviour quoted  in the preceding paragraph is  in sharp contrast
with the  situation encountered for equilibrium  diffusion models.  In
fact,  we find  that, as  a result  of the  use of  the  trace element
approximation, the  presence of mode trapping manifests  itself for all
of  the set  of  hydrogen envelopes  and  the whole  range of  periods
considered. This  is a  result that has  been found by  others authors
(Brassard et  al.  1992ab; Bradley 1996). As  mentioned, trapped modes
correspond  to local  minima  in  kinetic energy.  Note  that, as  \mh
decreases,  the number of  trapped modes  becomes markedly  lower, and
also that  the contrast  in kinetic energy  between trapped  modes and
other  (non-trapped) modes  increases,  as found  by  Brassard et  al.
(1992b)  (see  their figures  20a  and  20b).   As detailed  by  these
authors, for models with thick hydrogen envelopes, the hydrogen-helium
transition is located  at relatively large depths, in  such a way that
the radial  eigenfunctions of modes  adopt very low amplitudes.   As a
consequence,  the reduction  in  the amplitude  of the  eigenfunctions
caused by the  presence of the chemical transition  region is not very
appreciable for  the trapped modes, and therefore  their $E_{\rm kin}$
values  are not  very much  diminished  (see Fig.   4).  In
contrast, for  thin hydrogen envelopes, the lowering  of $E_{\rm kin}$
value   for   trapped   modes   is  more   noticeable,   because   the
hydrogen-helium  interface  is  located  in  a  region  in  which  the
eigenfunctions have larger amplitudes.

The situation is quite different when time-dependent element diffusion
is adopted  to modeling  the chemical interfaces.  As a result  of the
smoothness  caused by  element diffusion,  the eigenfunctions  are not
substantially    perturbed   by   the    density   changes    at   the
interfaces. Therefore, the modes have similar $E_{\rm kin}$ values, at
least from  a given $k$' value.  We  conclude that for all  of the \mh
values, the signatures of  mode trapping become therefore considerably
weaker   when  the   shape  of   the  chemical   composition   at  the
hydrogen-helium  transition   zone  is   assessed  in  the   frame  of
non-equilibrium  diffusion.    However,  note  that   the  differences
obtained between  both treatments of diffusion are  not very important
as  far as the  distribution of  $\Delta P_k$  is concerned.  A closer
inspection of the  behaviour of this quantity reveals  the imprints of
the other  chemical transitions, such as  the helium-carbon interface.
The signals of mode trapping  from this interface consist of secondary
minima in $\Delta P_k$.

Another interesting feature shown by these diagrams is the presence of
certain  modes  characterized  by  enhanced values  of  their  kinetic
energy.  The existence of these modes is related to the innermost peak
in the Ledoux term.  These modes are characterized by relatively large
amplitude in the  high-density, central region of the  star.  This can
be  appreciated  in  Fig.   6,  in  which we  depict  the  density  of
oscillation kinetic energy ($dE_{\rm kin}/dr$) in terms of the stellar
radius, corresponding to  modes with $k=$ 23, 24 and  25 and $\ell=$ 2
in a  model with  \mh= $-3.941$  (see Fig.  4).   As well  known, this
function is an indicator of how a given mode samples different regions
in  a stellar  model. This  function  has been  employed by  Gautschy,
Ludwig  \&  Freytag (1996).   Note  the  markedly  large amplitude  of
$dE_{\rm kin}/dr$  corresponding to the centrally  enhanced mode ($k$=
24),  as compared with  the neighbouring  ones ($k$=  23 and  25). The
presence of the centrally enhanced modes translates into very pronounced
minima in the $\Delta P_k$-$P_k$  diagrams. A similar finding has been
reported by Gautschy  et al. (1996).  It is also  worth noting that in
the case of stellar models with time dependent diffusion, the presence
of such  centrally enhanced  modes in the  spectrum of energy  is much
more apparent that in the  case of stellar models calculated under the
assumption   of   diffusive   equilibrium   in   the   trace   element
approximation,  in which case the imprint  of mode  trapping in  the outer
layers is the dominant feature.

\section{Conclusions} \label{sec:conclus}

In  this paper  we  have  explored the  effect  of the  time-dependent
element diffusion  on the mode  trapping properties of DA  white dwarf
models with various thickness  of the hydrogen envelope. Specifically,
we have computed adiabatic non-radial $g$-modes with $\ell =$ 1, 2, 3,
for periods ranging from 100 to  1000 s. In addition, we have obtained
the period spacing  and the oscillation kinetic energy  for each mode.
Pulsational  calculations have  been performed  on 0.6  $M_{\odot}$ WD
evolutionary  models in  a  $T_{\rm eff}$  range  covering the  entire
observed  ZZ Ceti  instability strip.   With respect  to  the hydrogen
content, the following values for the hydrogen envelopes \lmh= $-3.941$,
$-4.692$, $-5.672$, $-6.700$ and  $-7.349$ have been considered. The innermost
chemical composition of our models is that of Salaris et al. (1997).

Here, white dwarf modeling rests on a detailed and up-to-date physical
description.   In particular,  the processes  describing gravitational
settling, chemical and thermal  diffusion of nuclear species have been
considered in  the frame of  a multi-component treatment.  White dwarf
evolution  has  been  computed  in  a  self-consistent  way  with  the
predictions  of  element  diffusion   and  nuclear  burning.   In  the
interests of comparison, we have also performed pulsation calculations
in  the frame  of the  stellar models  constructed under  the standard
assumption   of   diffusive   equilibrium   in   the   trace   element
approximation.  In this sense, results using such a standard treatment
are in good agreement with those of other authors.

We find that time-dependent  element diffusion considerably smooth out
the composition  profile at the hydrogen-helium  chemical interface by
the time the  ZZ Ceti domain is  reached. This is true for  all of the
hydrogen envelopes we considered.  The main conclusion of this work is
that the  presence of mode  trapping becomes considerably  weaker when
the  shape   of  the  chemical  composition   at  the  hydrogen-helium
transition  zone is assessed  in the  frame of  time-dependent element
diffusion.    For  large  periods,   this  conclusion   remains  valid
regardless the hydrogen envelope mass with which a white dwarf settles
upon its  cooling track.  Notice  that such conclusion is  largely the
same we  have found in our  previous work on this  topic (C\'orsico et
al.  2001),  which was  based on white  dwarf models constructed  by a
full stellar evolution treatment. Let us remark that these models have
internal profiles somewhat different than those employed here (Salaris
et al. 1997).  Thus, it seems that, regardless the fine details of the
internal chemical  profile, the weakening  of mode trapping  caused by
time-dependent  element diffusion is  expected to  be correct  on very
general grounds.  Thus, we see no reason for considering mode trapping
as an efficient  mechanism to explain the scarcity  of eigenperiods in
the pulsational spectra of ZZ Ceti stars.

In closing,  we note  that in  view of the  smoothness of  the kinetic
energy distribution predicted by  our models with diffusively evolving
stratifications, the hypotetical existence of centrally enhanced modes
in real  stars would manifest itself  as the absence of  such modes in
the light curves.   This is so because these  modes are very energetic
and   would  have   low   probability  to   reach  sufficiently   high
amplitude. Thus, these modes (the absence of them in the light curves)
could be  a potentially  useful means for  exploring the  structure of
innermost deep regions of the WD cores.

\section{acknowledgments} 

We would like to acknowledge the comments and suggestions of an
anonymous referee, which improved the original version of this work.


\bsp
%

\begin{figure*}
\epsfxsize=450pt \epsfbox{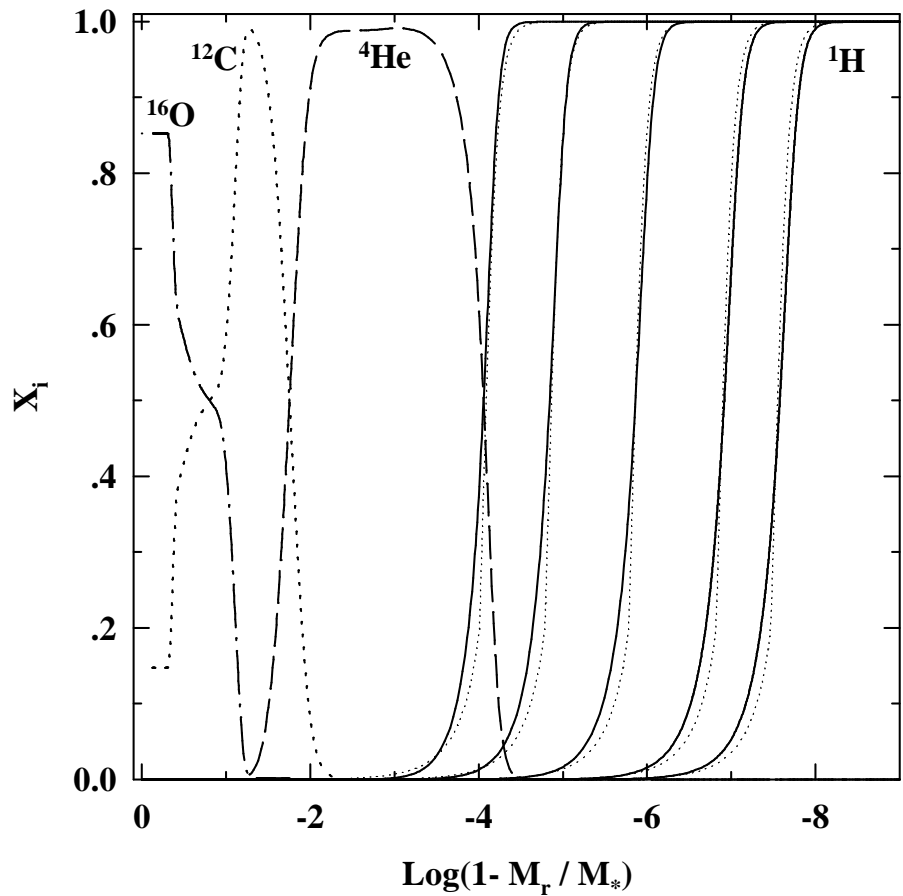}
\caption{The chemical abundance distribution of our stellar models for
all of the \mh values considered. Solid, dashed, dotted and dot-dashed
lines correspond to hydrogen,  helium, carbon and oxygen respectively.
In addition, hydrogen  profiles for models in which  the trace element
approximation is employed at the hydrogen-helium inteface are depicted
with thin dotted lines. The mass  of the models is 0.6 $M_{\odot}$ and
the effective  temperature is $\approx$  11800 K. For  clarity, except
for  the  thickest hydrogen  envelope,  the  helium  profile has  been
omitted.}
\end{figure*}

\begin{figure*}
\epsfxsize=450pt \epsfbox{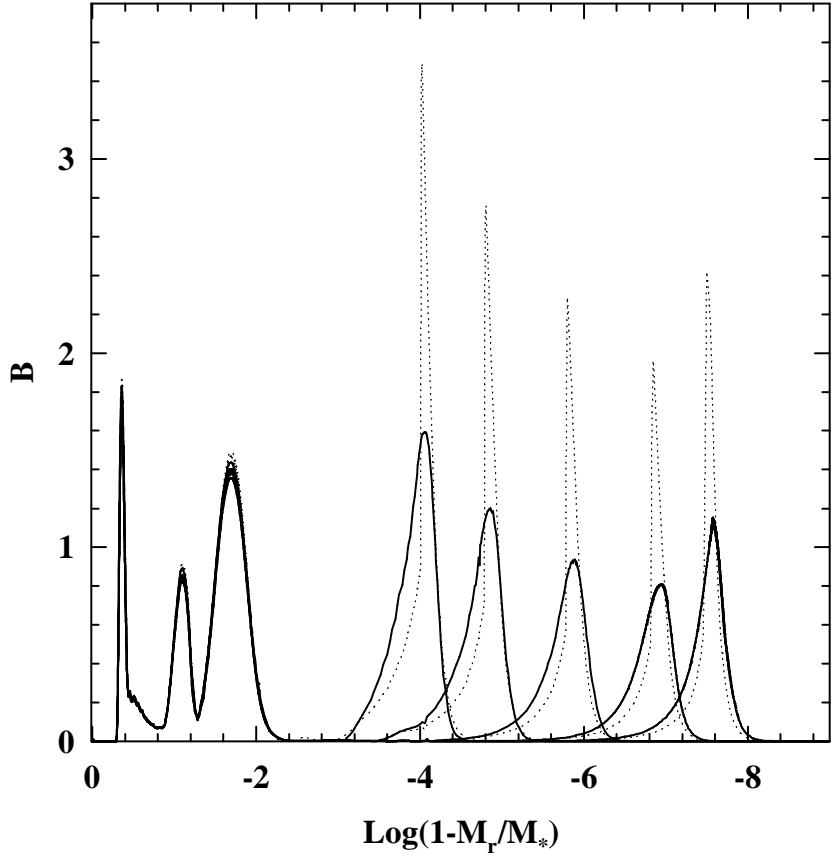}
\caption{The  Ledoux term  $B$ for  the same  models as  in  Figure 1.
Solid lines  correspond to our models  with non-equilibrium diffusion,
and  dotted  lines  depict   the  predictions  of  the  trace  element
approximation.  Note  that in  the latter case  $B$ shows  sharp peaks
with very  large values at the hydrogen-helium  interface, as compared
with  the  results  from  non-equilibrium diffusion.   An  outstanding
feature exhibited by  $B$ is the pronounced peak  at deepest layers of
the  models.   The  presence  of  such peak  is  responsible  for  the
existence of  some centrally enhanced modes in  the eigenspectrum. For
more details, see text.}
\end{figure*}

\begin{figure*}
\epsfxsize=450pt \epsfbox{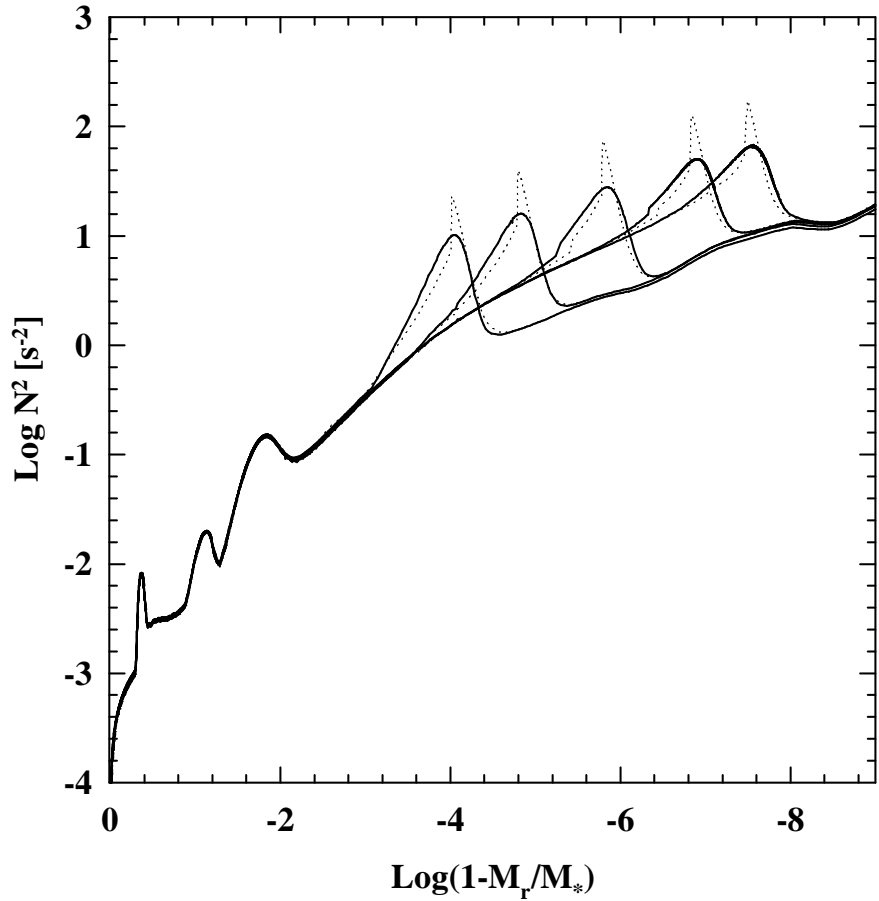}
\caption{Same  as figure  2, but  for the  squared Brunt-V\"ais\"al\"a
frequency.  Note  that for the  case of non-equilibrium  diffusion the
effect  of the  hydrogen-helium transition  is translated  into smooth
bumps, in  contrast to very sharp  peaks corresponding to  the case of
the trace element approximation.}
\end{figure*}

\begin{figure*}
\epsfxsize=450pt \epsfbox{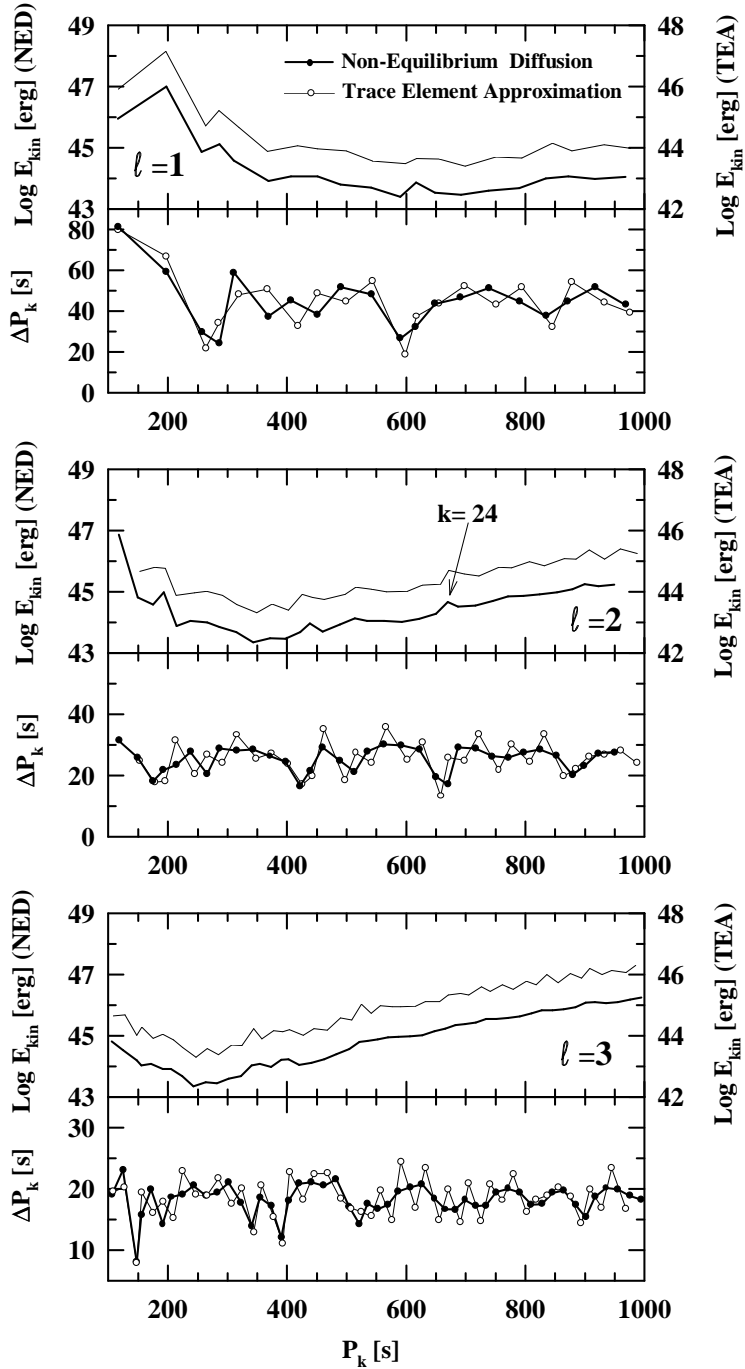}
\vskip -20mm
\caption{The  logarithm  of the  oscillation  kinetic  energy and  the
forward period  spacing for  harmonic degree $\ell$=1  (upper panels),
$\ell$=2  (centre  panels) and  $\ell$=3  (lower  panels),  for a  0.6
$M_{\odot}$  WD model  at  $T_{\rm  eff} \approx$  11800  K and  \lmh=
$-3.941$.  The values  of the  $E_{\rm kin}$  correspond to  the usual
normalization  $\delta r/  r=$  1 at  $r=  R_*$. The predictions of 
non-equilibrium diffusion (NED) are depicted by  filled  symbols and 
solid lines, whereas the results from the trace element  
approximation (TEA) are denoted by empty symbols and thin lines.
 With  the aim of remarking  
the differences in  the kinetic energy  behaviour  resulting from both  
treatments  of  diffusion,  symbols corresponding to eigenmodes have 
been  omitted in the $\log(E_{\rm kin})$-curves. In  the interests  of
clarity, the  scale for  the kinetic energy  in the case  of diffusive
equilibrium  is displaced  upwards by  1  dex. The  arrow indicates  a
centrally  enhanced  mode  with  $\ell=$  2, $k=  24$  (see  text  for
details).}
\end{figure*}

\begin{figure*}
\epsfxsize=450pt \epsfbox{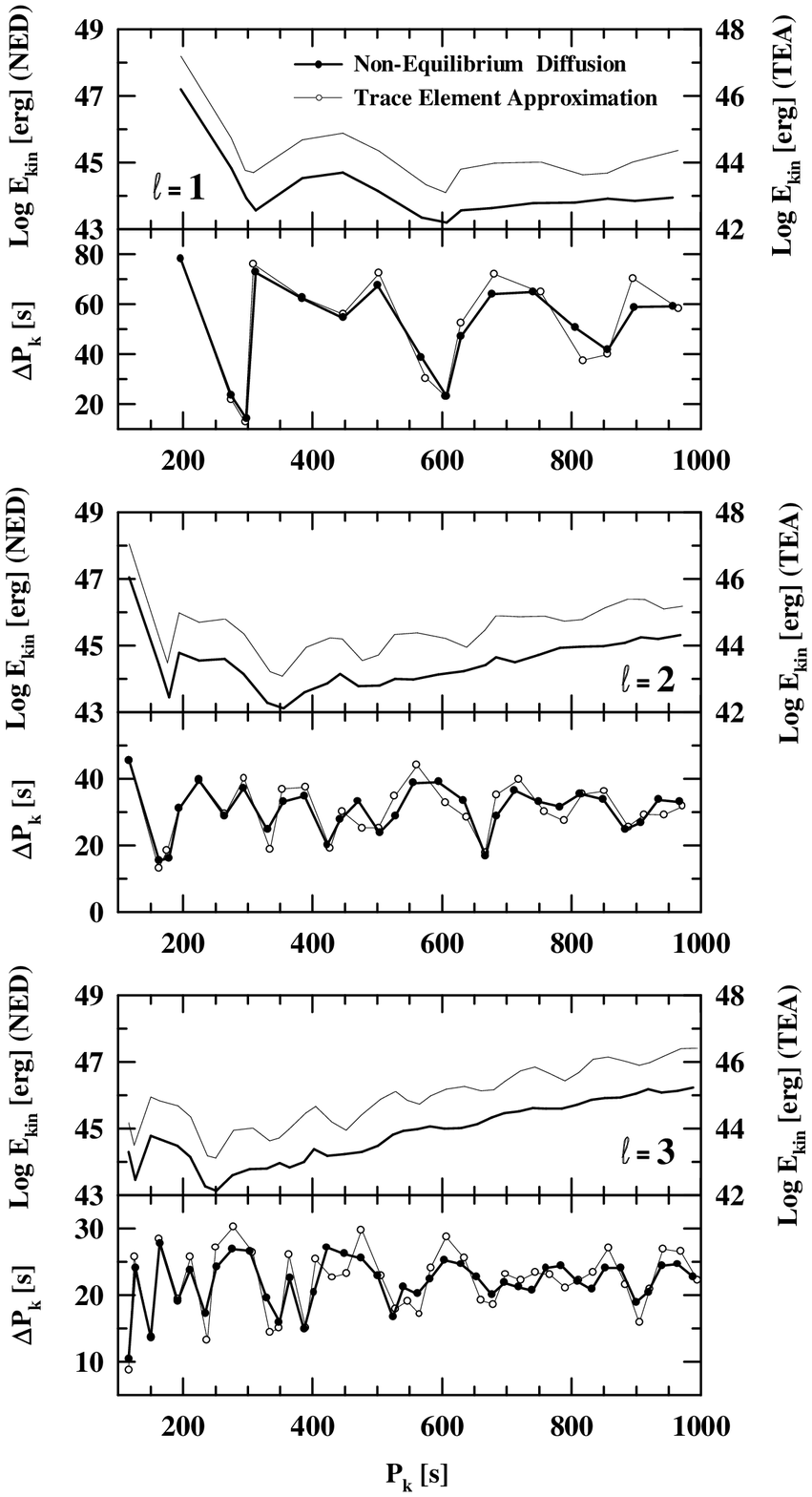}
\caption{Same as figure 4, but for  \lmh= $-7.349$.}
\end{figure*}

\begin{figure*}
\epsfxsize=450pt \epsfbox{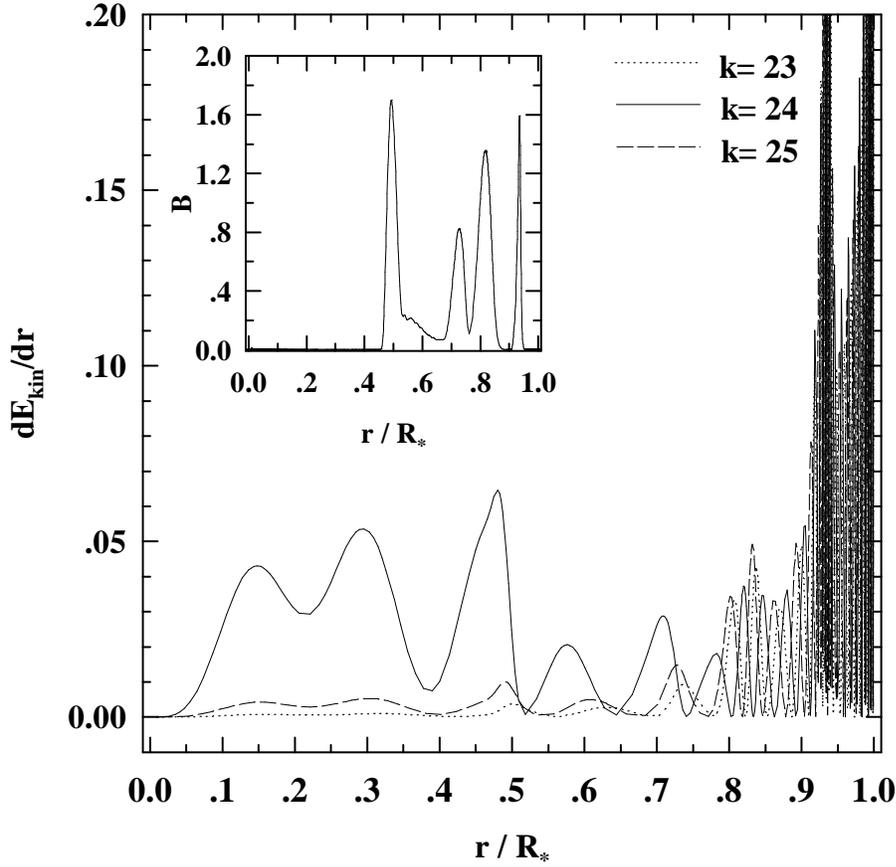}
\caption{The  density  of  the  oscillation kinetic  energy  ($dE_{\rm
kin}/dr$) in terms  of the stellar radius for  modes with radial order
$k= 23$ (dotted line), $k= 24$  (solid line) and $k= 25$ (dashed line)
for  the harmonic  degree $\ell=  2$. The  WD model  has  an effective
temperature of $T_{\rm  eff} \approx 11800$ K and  a hydrogen envelope
mass of \mh= $-3.941$. The inset  shows the profile of the Ledoux term
$B$.  Note that  the $k= 24$ mode (which is  a centrally enhanced one)
adopts  high amplitudes of  ($dE_{\rm kin}/dr$)  as compared  with the
adjacent modes.  This  is particularly true for the  region bounded by
the  stellar centre  and the  location  of the  first peak  in B  (see
inset).  As  a result, the $k= 24$  mode has an enhanced  value of its
oscillation kinetic energy ($E_{\rm kin}$) (see figure 4).}
\end{figure*}

\label{lastpage}

\end{document}